\documentclass[12pt,a4paper]{article}
\usepackage{mathtext}        

\usepackage{amssymb,amsfonts}
\usepackage[small]{caption2}
\usepackage{epsf}
\usepackage{rotate}
\parskip=\baselineskip

\makeatletter

\renewcommand{\section}{\@startsection
	{section}{1}%
        {0mm}%
        {1.5\baselineskip}%
        {0.5\baselineskip}%
        {\centering\normalfont\normalsize}}%

\renewcommand{\subsection}{\@startsection
	{subsection}{2}%
        {0mm}%
        {-\baselineskip}%
        {\baselineskip}%
        {\centering\itshape\normalsize}}%

\renewcommand{\@biblabel}[1]{#1.}

\makeatother
\newcommand{\Section}[1]{\section{\uppercase{#1}}}

\def\yr#1 {#1. }
\def\v#1 {V. #1. }
\def\p#1 {P. #1. }
\def\t#1 {Т. #1. }
\def\s#1 {С. #1. }

\begin{document}


\begin{center}
{\large\bf\uppercase{Neutron Stars in Globular Clusters: Formation and Observational 
Manifestations 
\\}}
{\bf
A. G. Kuranov* and K. A. Postnov** 
}
\end{center}

\begin{center}
{\it
Sternberg Astronomical Institute,  Moscow,  Russia \\
**E-mail: alex@xray.sai.msu.ru \\
***E-mail: pk@sai.msu.ru \\
}
published in Pis'ma v Astronomicheski Zhurnal, 2006, \\
Vol. 32, No. 6, pp. 438
\end{center}

\section{Abstract}

Population synthesis is used to model the number of neutron stars in globular clusters that are 
observed as low-mass X-ray sources and millisecond radio pulsars. The dynamical interactions between 
binary and single stars in a cluster are assumed to take place only with a continuously replenished 
''background'' of single stars whose properties keep track of the variations in parameters of the cluster 
as a whole and the evolution of single stars. We use the hypothesis that the neutron stars forming in binary 
systems from components with initial masses of $\sim 8-12 M_\odot$
 during the collapse of degenerate O-Ne-Mg 
cores through electron captures do not acquire a high space velocity. The remaining neutron stars (from 
single stars with masses $> 8 M_\odot$ or from binary components with masses  $>12 M_\odot$) are assumed to be 
born with high space velocities. According to this hypothesis, a sizeable fraction of the forming neutron 
stars remain in globular clusters (about 1000 stars in a cluster with a mass of $5\cdot 10^5 M_\odot$). The number of 
millisecond radio pulsars forming in such a cluster in the case of accretion-driven spinup in binary systems 
is found to be  10, in agreement with observations. Our modeling also reproduces the observed shape of 
the X-ray luminosity function for accreting neutron stars in binary systems with normal and degenerate 
components and the distribution of spin periods for millisecond pulsars. 

Key words: binary X-ray sources, neutron stars, millisecond pulsars, globular clusters. 

\section{INTRODUCTION }
Neutron stars (NSs) in globular clusters (GCs) 
are observed by currently available X-ray astronomical 
methods as accreting compact objects in lowmass 
X-ray binaries (including X-ray bursters) and 
by radio-astronomical methods as millisecond radio 
pulsars that are generally members of binary systems. 
There is an evolutionary connection between the 
two classes of objects - millisecond radio pulsars 
are accretion-driven spin-up products of old NSs 
in close low-mass X-ray binaries (this idea was 
first put forward by Bisnovatyi-Kogan and Komberg 
(1974); see also the review article by Bisnovatyi- 
Kogan (2006); for the most recent observational 
confirmations, see, e.g., Bogdanov et al. (2005)). The 
standard scenario for the evolution of low-mass Xray 
binaries and millisecond pulsars (see, e.g., Bhattacharya 
and van den Heuvel 1991) in dense GCs is 
complemented by the effect of dynamical interactions 
between stars - new binaries formed through tidal 
interactions, captures, component exchanges, etc. 
are added to the originally existing binary systems. 

Observational data clearly show an increased concentration 
of X-ray sources and binary millisecond 
pulsars in GCs (see, e.g., Heinke et al. 2005; Grindlay 
2004; Lorimer 2005), thereby confirming that dynamical 
effects are crucially important for describing 
their formation and evolution (Pooley et al.2003). 

NSs are formed through the collapse of the cores 
of stars with main-sequence masses above a certain 
value 8-10 $M_\odot$, determined by the chemical composition 
and, possibly, other parameters, that finished 
their thermonuclear evolution. Depending on the evolutionary 
status of the star at the time of its collapse, 
the NS birth through core collapse is accompanied 
by a type-II or type-Ibc supernova explosion and by 
the ejection of a significant presupernova mass in 
the form of a shell. Data on young NSs observed as 
radio pulsars are also indicative of their high space 
velocities with amean value of $\sim 400$ km~s$^{-1}$ (see, e.g., 
Lorimer (2005) and references therein; Hobbs et al. 
2005). The hypothesis about a kick velocity acquired 
by a NS during collapse is commonly invoked to 
account for such high space velocities. The physical 
mechanism underlying this velocity is unknown 
and there are many speculations and models on this 
score; many of them were listed and discussed in the 
review by Lai (2001). Note that deducing the shape of 
the distribution of kicks from observations is coupled 
with difficult allowance for selection effects (Tutukov 
et al. 1984) and it is believed (see, e.g., Iben and 
Tutukov 1996) that nature does without kicks and 
all of the high space velocities of radio pulsars are 
attributable to their birth in close binary systems. 

The hypothesis of high space velocities of young 
NSs immediately leads to the well-known problem of 
NS retention in GCs, since the escape velocity even 
in the densest clusters does not exceed several tens 
of km/s. Most of the NSs born through the core 
collapse of massive stars should have escaped from 
the cluster shortly after its formation. 

In principle, the problem of NSs in GCs can be 
solved by assuming a nonstandard mechanism of 
their formation in GCs, for example, through the collapse 
of white dwarfs in close binary systems induced 
by mass accretion from the secondary component 
(Helfand et al.1983; van den Heuvel 1984; Grindlay 
and Bailyn 1988; Bailyn and Grindlay 1990). In this 
scenario, the white dwarf must be composed of a mixture 
of O, Ne, and Mg (Miyaji et al.1980); mass accretion 
onto a carbon-oxygen white dwarf ends with 
its total destruction during the explosive thermonuclear 
burning that arises when its mass approaches 
the Chandrasekhar limit, producing a type-Ia supernova 
(for recent calculations, see, e.g., Dunina- 
Barkovskaya et al.(2001) and references therein). 

Recent evidence suggests that, possibly, not all 
of the young NSs are born with anomalously high 
space velocities. In particular, this conclusion is 
reached when analyzing evolutionary scenarios that 
give rise to binary NSs like the binary radio pulsar 
PSR J0737-3039 (Bisnovatyi-Kogan and Tutukov 
2004; Dewi et al.2005; Podsiadlowski et al.2005). In 
this binary, the mass of the young pulsar is 1.25 $M_\odot$, 
the mass of the old (spun up by accretion) millisecond 
pulsar is 1.34 $M_\odot$, and the orbital eccentricity is 
low (e = 0.088). This is indicative of a low mass 
ejected during the explosion of the second supernova 
in this binary and a low kick velocity acquired by the 
NS during collapse. It was surmised that the NSs 
formed through the collapse of an O-Ne-Mg stellar 
core triggered by electron captures do not acquire 
kick velocities if these stars are members of binary 
systems (Podsiadlowski et al.2004; van den Heuvel 
2004). In all of the remaining cases (a single star with 
$M>8 M_\odot$ or a star with a mass $>12-14 M_\odot$ in 
a binary), the stellar core collapses anisotropically 
and the forming NSs will acquire significant space 
velocities. As follows from currently available detailed 
calculations of the collapse of O-Ne-Mg stellar cores 
induced by electron captures (Kitaura et al.2005), the 
mass of the forming NS is 1.25 $M_\odot$. 

Note that the range of initial masses for mainsequence 
stars in which degenerate O-Ne-Mg cores 
are formed is not completely clear. In the first calculations 
(Miyaji et al.1980), this range was 8-12 $M_\odot$, 
but Iben with coauthors (Ritossa et al.1999) subsequently 
narrowed it to 10-11 $M_\odot$; other evolutionary 
calculations (Pols et al.1998) yield the range 
5-8 $M_\odot$, depending on the chemical composition. Obviously, 
the particular features of evolutionary calculations 
and, possibly, the rotation of stars are important 
here. In our calculations, we assumed this range to 
be 8-12 $M_\odot$, since the exact width of this range has 
no effect on the final results due to the presence of a 
much less certain parameter, the initial binary fraction 
in GCs. 

Simple estimation shows that for the assumed 
Salpeter initial mass function $f(M)\sim M^{-2 .35}$ with 
a minimum mass of $0.1 M_\odot$ and 100\% of binaries in a 
GC with a mass of $5\cdot 10^5 M_\odot$, the expected number 
of NSs in the cluster, according to the hypothesis 
under consideration, is $\sim 1000$. This number is by 
two orders of magnitude larger than the expected 
number of NSs under the assumption of a universal 
kick velocity with a mean value of 400 km/s for all 
NSs, which can be obtained from measurements of 
the space velocities for single radio pulsars (Hobbs 
et al.2005). 

Clearly, whether the hypothesis about two types of 
NSs is valid can be established only after elucidating 
the physical cause of the possible collapse anisotropy. 
However, it seems of interest check whether this hypothesis 
agrees with observational data on the number 
ofNSs of different types in GCs, since the absence 
of a kick velocity during collapse in some of the NSs 
automatically helps solve the problem of their retention 
in GCs without invoking other assumptions. 
This was first pointed out by Pfahl et al.(2002) and 
Podsiadlowski et al.(2004). 

We used an approach to calculating the evolution 
of compact stars in GCs based on the population 
synthesis method in combination with allowance for 
the dynamical interactions between stars in dense 
star clusters developed at the Sternberg Astronomical 
Institute of theMoscow State University (see Kuranov 
and Postnov 2004). In this paper, we show that 
the hypothesis under consideration in combination 
with the universally accepted assumption ofNS magnetic 
field decay at the accretion phase (Bisnovatyi- 
Kogan and Komberg 1974) successfully explains the 
observed properties of NSs in GCs, including their 
number and distributions in X-ray luminosity (for 
accreting NSs in low-mass X-ray binaries) and spin 
period (for millisecond pulsars). 

\section{THE MODEL }

The fundamental difficulty in describing the evolution 
of dense star clusters lies in the necessity 
of taking into account the mutual influence of the 
evolution of the stellar population on the evolution 
of the cluster as a whole, on the one hand, and 
the reverse effect of the change in cluster parameters 
on the rate of dynamical interactions between 
cluster-populating stars, on the other hand. The 
currently available population synthesis codes allow 
the evolution of single and binary stars in the Galaxy 
to be modeled. There are also codes that allow 
the evolution of self-gravitating systems consisting 
of a large number of star ($\sim 10^6$) either of equal 
masses or in a limited mass range to be computed. 
At present, intensive work is being done to modify 
and combine the existing codes into a single 
code. The international working group MODEST 
(http://www.manybody.org/modest/) was created in 
recent years to solve this problem and substantial 
progress has been achieved in the above field owing 
to its activity. 

The code presented here implements one of the 
possible approaches to describing and modeling the 
evolution of binary stars in GCs without resorting 
to direct N-body calculations. The basic method of 
calculations consists in the following. A newly formed 
binary system (an initial binary or a binary formed 
during the GC evolution as a result of dynamical 
interactions) is placed in a GC where it evolves with 
allowance made for its interactions with the surrounding 
stars. The evolution of the GC and ''background'' 
stars is assumed to be known, is identical for 
each computed binary, and does not depend on the 
evolution of this binary in the GC model used. The 
evolutionary track of the binary consists of a finite 
number of phases. The phases change either through 
the stellar evolution of one of the components (the 
evolution of an ''isolated'' binary) or through dynamical 
interactions with the surrounding stars. Our code 
includes the following types of interactions: 
\begin{enumerate}
\item
 The interactions that lead to changes in the 
orbital parameters of binaries (semimajor axis and 
eccentricity) - flyby; 
\item The close passages that lead to star exchanges; 
\item The passages that lead to the disruption of 
binary systems; 
\item The interaction between single stars that lead 
to the formation of binary systems (tidal captures). 
\end{enumerate}
Thus, we assume that the evolution of the cluster 
and its stellar population affects the evolution of the 
computed binary, but we disregard the reverse effect 
of this binary on the evolution of the cluster and the 
surrounding stars. This approach is somewhat limited 
(in particularly, it does not allow the interactions between 
binary systems and the interactions with rare 
objects, e.g., black holes, to be taken into account), 
but it has a number of advantages. These include the 
following: 
\begin{enumerate}
\item A fairly high computational speed; 
\item The ''clarity'' of calculations that allows the 
degree of influence of particular processes on the evolution 
of binaries to be estimated; 
\item The possibility of calculating an arbitrary (statistically 
significant) number of evolutionary tracks 
for binary stars within the framework of a single GC 
model. 
\end{enumerate}

\subsection{Initial Distributions and the Evolution of a Binary 
System in a GC }
An improved version of the scenario machine population 
synthesis code (Lipunov et al.1996) adapted 
to calculate the evolution in dense stellar systems is 
used to model the detailed evolution of binary systems.
We use the following standard initial distributions: 
\begin{equation}
\label{loga}
f(\log{a})=const, ~~~~~~~~~~ max \left\{
\begin{array}{lcl}10R_\odot\\
\rm{R_L(M_1)}\\
\end{array}
\right\}
<a<10^3  R_\odot,
\end{equation}
for the semimajor axes of binary stars and 
\begin{equation}
\label{m1}
f(M_1)~\propto M^{-2.35}_{1},~0.1M_\odot<M_1<120M_\odot\\
\end{equation}
for the mass of the primary component. Here, $R_L(M1)$ 
is the Roche-lobe radius for the primary component. 
For the component mass ratio, we use a power-law 
distribution: 
\begin{equation}
\label{m2}
f(q)~\propto q^{\alpha_q},~q=M_2/M_1<1.
\end{equation}
Here, $\alpha_q$ is the parameter of the distribution in component 
mass ratio. In our calculations presented below, 
we set $\alpha_q=0$, i.e., we considered a ''flat'' distribution 
in component mass ratio. 

The dipole surface magnetic field $B$ of the forming 
neutron stars was assumed to have a uniform $\log$ 
distribution: 
\begin{equation}
  f(B)\,dB \propto d\log B, \quad 10^{10}\hbox{G} <B< 10^{14} \hbox{G}
\label{e:initial:mu-ravn}
\end{equation}
which spans the range of field strengths estimated 
from observable single radio pulsars (except millisecond 
pulsars) in the absence of magnetic field decay. 
The magnetic field was assumed to decay only 
for NSs in binary systems at the accretion phase. 

In this case, we used an exponential law of magnetic 
field decay with time up to a certain lower limit 
(van den Heuvel et al.1986): 
\begin{equation}
B=\left\{
\begin{array}{lr}
        	B_0 e^{-t/\tau}\,,\quad & B > B_{\min}\\
                B_{\min}\,     & t > \tau\,\ln(B_0/B_{\min})\\
\end{array}
\right .
\label{e:compact:mu(t)}        
\end{equation}

We took the field decay time scale to be $\tau=10^7$ yr and 
the minimum magnetic field strength to be $B_{min}\simeq10^8$ G. 

As we noted in the Introduction, NSs acquire a 
kick velocity during their birth. This velocity was 
modeled as follows: we assumed the direction of 
the kick velocity to be random and isotropically 
distributed and determined its magnitude as 
\begin{equation}
f(v)\,dv~~ \left\{
\begin{array}{lr}
                =0, \quad &8 M_{\odot}<M<12M_{\odot}  \\
        	\propto \exp\left(-\frac{v^2}{2w^2}\right)\,v^2\,dv, & M>12M_{\odot}  \\
\end{array}
\right .
\end{equation}
for NSs from binaries and as 
$$
f(v)\,dv~~\propto \exp\left(-\frac{v^2}{2w^2}\right)\,v^2\,dv 
$$
for NSs from single stars. In both cases, the distribution 
parameter $w$ was set equal to $250$ km/s, in close 
agreement with the velocity measurements of single 
radio pulsars (Hobbs et al.2005). 

The initial NS spin periods were assumed to be 
identical for all stars and equal to $P_0=10$ ms. 

\subsection{Background Stars in GCs }
To properly describe the dynamical interactions 
of a binary system with other GC objects, the code 
also computes the evolution of single GC stars. The 
problem is simplified, because information about the 
distribution of stars only in mass (and radius, for tidal 
captures) at each time will suffice to determine the 
rates of dynamical interaction. All of the remaining 
parameters of a single star at the current time will be 
required only if it becomes a binary component as a 
result of the exchange interaction or the tidal formation 
of a new binary. This allows us to describe the 
background of single GC stars by creating a discrete 
set $N(m_i,R_j,t)$, where $m$ and $R$ are, respectively, the 
mass and radius of the star at time t, so that 
$$N^t_{Total}=\sum _{i}^{n_m}\sum _{j}^{n_r} N(m_i,R_j,t) ,$$
where $N^t_{Total}$ is the total number of single GC stars at 
time t. In our calculations, we assumed that nm = 45 
and nr = 45. The following processes were taken into 
account in the formation of the background of single 
stars: 
\begin{enumerate}
\item
The evolution of single stars. 
We assumed that the evolution of a single star 
was entirely determined by its mass (for a given 
chemical composition): $M_i(t)=f(M_i^0,t)$. Choosing 
a star from the mass interval $(M_i(t),M_i(t)+\delta m_i)$ at 
time $t$ and using the dependence $M_i^0=f^{-1}(M_i(t))$, 
we then randomly draw the initial mass of the star, 
according to Eq.(2), from the initial interval $(M_i^0,M_i^0+\delta m_i^0)$.
 In turn, knowledge of the initial mass 
allows us to calculate the evolution of the single 
star under consideration and to determine all of the 
remaining required parameters at time $t$. 
Thus, the set $N(m_i,R_j,t)$ allows the stellar population 
of single GC stars to be completely described. 

\item The space distribution of background stars.
 
Mass segregation results in the settling ofmassive 
stars to the GC center.W e used the Michie-King 
multimass model (Michie 1963) to describe the GC 
structure. The space density  $\rho_i$ of a stellar subsystem 
formed by stars with masses in the mass range 
$(M_i(t),M_i(t)+\delta m_i(t))$ (corresponding to the breakdown 
of the set $N(m_i,R_j,t)$) is described by a power 
law: 
\begin{equation}
\rho _i(r) =
\left\{
\begin{array}{lr}
\rho _{c_i}, &r\le r_{c_i}\\
\rho _{c_i}(r/r_{c_i})^{-2}, &r_{c_i} <  r\le r_{h_i}\\
\rho _{h_i}(r/r_{h_i})^{-4}, &r_{h_i} <  r\le r_t\\
\end{array}
\right.
\label{rho}
\end{equation}
where $r_{c_{i}}$ and $r_{h_i}$ are, respectively, the radii of the 
core and the sphere in which half of the mass of the 
stellar subsystem is contained, and $r_t$ is the tidal 
radius of the cluster. Thus, $\rho_{h_i}=\rho_{c_i}(r_{h_i}/r_{c_i})^{-2}$ and 
$\rho_{c_i}=M_{tot_i}/[8\pi r_{c_i}^2(r_{h_i}-\frac{2}{3}r_{c_i})] $. 
In the case of energy 
equipartition for GC stars, we may assume that 
$$r_{c_{i}}=\sqrt{\frac{\bar{m}_c}{m_i}} r_c, $$
where $\bar{m}_c$ is the mean mass of the stars in the GC 
core, $r_c=\sqrt{\frac{3v_m(0)^2}{4\pi G \rho_c}}$
is the GC core radius, $v_m$ is the 
rms space velocity, and $\rho_c$ is the central star number 
density: 
\begin{equation}
\rho_c=\sum_i \rho_{c_i}
\end{equation}

\item The GC evolution.

In general, the distribution 
$\rho_i(r)$ is a function of time, reflecting the GC evolution. 
In this formulation of the problem, the GC evolution 
is important from the viewpoint of the change in the 
rate of interaction between stars. To take into account 
the GC evolution (the collapse of the cluster core followed 
by its expansion), we used a power law for the 
changes in parameters $r_c$ and $v_m(t)$, as follows from 
numerical calculations (see, e.g., Kim et al. 1998). 
The core collapse at late phases is described by the 
formulas: 
\begin{equation}
\label{r1(t)}
\rho_c(t)\sim~\rho_c(0)~(1-t/t_{coll})^{-1.2},
v_m(t)\sim~v_m(0)~(1-t/t_{coll})^{-0.12},
\end{equation}
where $t$ is the current time and $t_{coll}$ is the cluster core 
collapse time. 

As our numerous calculations show, the core collapse 
stops as a result of the formation of binary stars 
in the central part of the cluster. The succeeding core 
expansion can also be described by a power law: 
\begin{equation} 
\rho_c(t)\sim t^{-2},
v_m(t)\sim t^{-0.32}.
\label{v2(t)}
\end{equation}

The changes in parameters $\rho_c$, $r_c$ and $v_m$ during the 
cluster evolution are schematically shown in Fig.1. 
We used two GC models with different central densities 
$\rho_c$ and core radii $r_c$ at the collapse time (see the 
table and Fig.1). In both models, the collapse time 
was taken to be $t_{coll}=7\cdot 10^9$ yr. At the GC formation 
time, the mass of all cluster stars was set equal to 
$5\cdot 10^5 M_\odot$; half of all cluster stars were assumed to 
be binary. Roughly speaking, one third and two thirds 
of the initial cluster mass are concentrated in single 
and binary stars, respectively. 

\item
The evaporation of stars from the GC.

The set $N^t(m_i,r_j)$ also changes through the evaporation 
of background stars from the GC and the formation/
disruption of binary systems: 
$$
N(m_i,r_j,t)=N(m_i,r_j,t-1)
-\Delta N_{evap}^{t}(m_i,r_j)
- \sum _{k,l}
\Delta N_{bs}^{\delta t}(m_i,r_j,m_k,r_l)),
$$
where $N(m_i,r_j,t)$ is the number of GC binaries in 
the corresponding phase volume element $(\delta m_i,\delta r_j)$,
$\Delta N_{evap}^{t}(m_i,r_j)$  is the change in $N(m_i,r_j,t)$  through 
the evaporation of stars from the cluster core, and 
the last term describes the decrease in the number of 
single stars through the formation of tidal binaries. In 
this case, we assumed that 
$$
\Delta N_{evap}^{t-1}(m_i,r_j)=N^{t-1}(m_i,r_j) c_{evap}(m_i),
$$
where the coefficients $c_{evap}(m_i)\sim m_i^{-1}$. Lighter stars 
with lower masses mostly go into the halo and their 
cluster binding energy decreases. This behavior is 
to be expected from the main properties of the slow 
relaxation via pair collisions. The time scale for this 
process was estimated by Spitzer (1969) (see also 
Mouri and Taniguchi 2002): 
$$t_s=\frac{(v_1^2+v_2^2)^{3/2}}{8(6\pi)^{1/2}G^2
m_1m_2n_1\ln(\Lambda)}=1.2\frac{m_1}{m_2}t_{r0}(m_1).$$
\end{enumerate}

\subsection{ Interactions between GC Stars }
The orbital parameters of a binary system moving 
inside a GC are determined by its energy $E$ and angular 
momentum $J$. The relative motions of GC stars 
produce continuous fluctuations in the gravitational 
field; in turn, these fluctuations cause the quantities $E$ 
and $J$ to change. The energy and angular momentum 
can also change through close encounters and if a 
supernova explosion occurred during the evolution of 
the massive binary component, as a result of which 
the binary instantaneously loses its mass and can 
acquire an additional kick due to collapse anisotropy. 
When each evolutionary track of a binary system is 
calculated, the changes in $E$ and $J$ with time are 
tracked; for a detailed description, see our previous 
paper (Kuranov and Postnov 2004). 
The rate of interaction between a binary and a 
subsystem of stars $i$ with a number density $n_i(r) = \rho_i(r)/ m_i$ at time $t$ is determined by the orbit-averaged 
value 
$$
{\cal R}_{ij}(t)=\sigma_{ij}\times \frac{2}{P(E,J)} \int_{r_p}^{r_a} 
n_i(r,t)v(r,t)dr/v_r\,, 
$$ 
where $r_p$ and $r_a$ are, respectively, the orbital pericenter 
and apocenter of the star in the cluster, $v$ is the space 
velocity of the binary, $v_r$ is the radial component, $P$ is 
the orbital period, 
$$ P(E,J)= 2\int_{r_p}^{r_a} dr/v_r.$$
The subscript j(j = 1. . . 4) corresponds to the type of 
dynamical interaction between this binary and single 
stars and $\sigma_{ij}$ is the cross section for the corresponding 
process. 
The probability of a dynamical interaction between 
the binary under consideration and single background 
stars during time .dyn is defined as 
\begin{equation}
p(\tau_{dyn})= 1-exp{\int_{ts}^{ts+\tau_{dyn}} {\cal R}(t)dt },
\label{time}
\end{equation}
where the total rate of possible interactions is 
$$ 
{\cal R}(t)=\sum_{i,j}{\cal R}_{ij}(t)\,.
$$
The scenario machine population synthesis method 
uses a discrete description of the evolution of binary 
components. In a GC, the evolutionary phase of a 
binary changes in time $dt=min\{dt_{evol},\tau_{dyn}\}$, i. e. , either 
through the intrinsic evolution of one of the components 
(the evolution of a binary ''isolated'' from the 
background) or through its dynamical interactions 
with the surrounding stars. The time $dt_{evol}$ depends 
only on the state of the binary components at the 
beginning of each phase and $\tau_{dyn}$ can be determined 
from Eq.(\ref{time}), in which the probability of a dynamical 
interaction $p(\tau_{dyn})$ is specified by a random variable 
uniformly distributed in the interval $[0, 1]$. 

To choose the type of interaction $j$ and subsystem $i$ 
of single background stars involved in the interaction, 
let us introduce quantities $p_{ij}$ that have the meaning 
of the probability of interaction $j$ with subsystem $i$ of 
single background stars for the binary system under 
consideration: 
$$
p_{ij}= {\cal R}_{ij}/{\cal R}, ~~~ \sum_{i,j}{p_{ij}}=1.
$$
Next, we choose a random number p uniformly distributed 
in the interval $[0, 1]$. We will sumthe elements 
of matrix $p_{ij}$ (whether over its rows or columns) until 
the sum will exceed the number $p$. The element $i$, $j$ in 
which this occurs will define the sought-for values of $i$ 
and $j$. This procedure ensures an appropriate choice 
of indices $i$, $j$. This procedure is illustrated particularly 
clearly when one or more elements dominate in the 
matrix $p_{ij}$ and when most of the matrix elements are 
of the same order of magnitude. 
\Section{RESULTS OF CALCULATIONS }
The results of our calculations for the evolution of 
binary systems in GCs for models A and B are shown 
in Figs.2 -8. To obtain statistically significant results 
for each GC model, we computed the evolution of one 
million binary systems by the Monte Carlo method. 
The background of single stars with which the binary 
system interacts was computed in advance for each 
GC model. The neutron stars that did not escape from 
the cluster, including those formed in binary systems 
from stars in the mass range 8-12 $M_\odot$ without any 
initial kick velocity, were included in this background. 

In Fig.2, the total number of binaries with neutron 
stars is plotted against time with a breakdown into 
subtypes NSs in pairs with main-sequence stars, 
NSs in pairs with white dwarfs, binary NSs, and 
NS-black hole pairs. The binaries retained in the 
GC are shown in the left panels. The binaries that 
escaped from the GC are shown in the right panels. 
The dynamical GC core collapse time is marked by 
the vertical dotted line. As might be expected, the 
number of binaries increases after core collapse. The 
decrease in the number of binaries with time for some 
of the binary types results from component mergers 
(this is particularly clearly seen for the binaries that 
escaped from the cluster) and dynamical disruptions. 
However, the latter effect causes no appreciable decrease 
in the total number of binary stars inside the 
cluster, but more likely manifests itself as a slowdown 
in the increase in the number of stars with time. The 
difference in the models of the GC itself is more pronounced 
for the binaries that escaped from the cluster: 
in the more centrally condensed cluster B (lower panels), 
the number of binaries escaping from the cluster 
increases due to more intense dynamical interactions 
in the cluster core. Note also that the number of 
produced binary NSs proved to be very small; to 
obtain a statistically significant and proper estimate 
of the formation rate of these important astrophysical 
sources in GCs requires calculating a larger number 
of stars and taking into account the collisions with 
background binary stars (Grindlay et al.2005). We 
plan to perform such calculations in the future. 

Figure 3 shows the formation rate of binary systems 
with neutron stars in the case of dynamical 
interactions between stars inside the GC: exchange 
interactions (an impinging single NS occupies the 
place of one of the binary components) and the formation 
of binary systems through tidal captures. The 
formation rate through exchange interactions is considerably 
higher that the formation rate through tidal 
captures for both GC models. In the overwhelming 
majority of cases, a main-sequence star is the NS 
companion for exchange interactions (the formation 
rate of binary systems with a white dwarf as the secondary 
component does not exceed 1Gyr$^{-1}$ and is not 
shown in the figure). The formation rates of accreting 
NS observed as low-mass X-ray binaries (LMXBs) 
and millisecond pulsars (MPSRs) formed through 
the same dynamical interactions between stars in the 
GC (exchange interactions and tidal captures) are 
shown in Fig.4. 

Figure 5 shows the time evolution of the total 
number of low-mass X-ray binaries and millisecond 
radio pulsars. Different shading indicates the change 
in the number of close binary systems in which accretion 
onto the NS takes place when a main-sequence 
dwarf and a degenerate low-mass white dwarf fill 
their Roche lobes due to the loss of orbital angular 
momentum through gravitational waves. We see that 
the number of objects of this type escaped from the 
cluster is comparable to the number of objects left 
inside the cluster and depends weakly on the chosen 
cluster model. The ratio of the number of millisecond 
radio pulsars to the number of low-mass Xray 
binaries is shown separately in Fig.6. Note that 
the number of millisecond pulsars spun up by accretion 
in the computed models is approximately half 
the number of accreting NSs. One would think that 
this situation is inconsistent with the observations 
(recall that one or two strong X-ray sources, many 
weak X-ray sources, and 10-20 millisecond pulsars 
are observed in a typical GC). However , including 
the initially single NSs at the ejection phase (radio 
pulsars) changes the situation: the ratio of the total 
number of NSs observed as radio pulsars (spun-up 
pulsars + NSs with initially weak magnetic fields) to 
the number of accreting NS is larger than one (the 
right panel in Fig.6). It is theoretically clear that the 
initial distributions of NSs in period and magnetic 
field and the evolution of the NS magnetic field are 
crucially important in analyzing these ratios. As regards 
the observed situation with X-ray sources, as 
we noted above, the situation with weak unidentified 
X-ray sources remains uncertain and the number of 
quiescent LMXBs among the weak X-ray sources in 
GCs can increase with time. 

Figure 7 shows the time evolution of the X-ray 
luminosity function for GC. We see that the luminosity 
function cannot be described by a single power 
law. The smooth steepening of this function near 
$10^{37}$ erg s$^{-1}$ may be due to the difference in the 
regimes of accretion onto the neutron star (Postnov 
and Kuranov 2005). The mean slope of the X-ray 
luminosity function, $d\log N/d\log L_x\sim -1.3$, is close 
to the observed values of the luminosity function for 
low-mass X-ray binaries in galaxies (Fabbiano 2005) 
before the knee at $\sim
5\cdot 10^{38}$ erg s$^{-1}$. In our simplified 
calculations, we could not obtain stationary X-ray 
sources with luminosities above the Eddington limit 
in the case of accretion onto the neutron star. 

The period distribution for millisecond pulsars is 
shown in Fig.8. Recall that the initial spin period 
of all NSs in our calculations was $P_0 = 10$ ms 
(the second peak in the figure corresponds to this 
value). The presence of pulsars with periods shorter 
than $P_0$ implies that these stars have passed the phase 
of accretion-driven spin-up in binary systems. We 
clearly see that the fraction of these NSs increases 
monotonically after the GC core collapse stops. The 
derived peak at millisecond periods is in qualitative 
agreement with the observed period distribution of 
pulsars in GCs (for the GCs Terzan 5 and 47 Tuc, see 
Fig.3 from Camillo and Rasio 2005). 

\Section{DISCUSSION AND CONCLUSIONS }
Here, we presented the results of our calculations 
of the formation and evolution of neutron stars in 
dense globular clusters. Our simulations were performed 
by the population synthesis of the evolution 
of binary and single stars in the approximation of a 
persistent background of single stars with which they 
interact dynamically. We considered two GC models 
with different star number densities and different core 
sizes after dynamical collapse with given time evolution 
of parameters. 

We assumed that the NSs born in binary systems 
through the collapse of degenerate O-Ne-Mg stellar 
cores induced by electron captures (this was thought 
to be possible at the end of the evolution of stars 
with initial main-sequence masses in the range 8-12 $M_\odot$ do not acquire any kick velocity. In all of 
the remaining cases of NS formation (i.e., from single 
stars with masses $>8 M_\odot$ or from binary components 
with masses $>12 M_\odot$), the newly born NSs were 
assumed to acquire a kick velocity in accordance 
with a Maxwellian distribution with a parameter of 
$250$ km s$^{-1}$. In this scenario, a significant fraction 
of the forming NSs is retained in the cluster. About 
1000 NSs remain in a cluster with a mass of $5\cdot 10^5 M_\odot$ in which half of its stars were initially binary 
members. In this case, the initial masses of such NSs 
must be 1.25 $M_\odot$. Observations of the GC 47 Tuc 
(Heinke et al.2005) show that the space distribution 
of millisecond pulsars in this cluster corresponds to 
a mean NS mass of $\sim 1.39 M_\odot$. Since $\sim 0.1 M_\odot$  is 
required to spin up NSs to millisecond periods at the 
accretion phase (Lipunov and Postnov 1984), these 
measurements can serve as an indication of an initial 
NS mass of $1.3 M_\odot$. 

Our modeling of the number ofmillisecond pulsars 
spun up by accretion in binary systems showed good 
agreement with the observational data: for the cluster 
models under consideration, we obtained  $\sim5-10$ millisecond pulsars and approximately the same 
number of such objects are dynamically ejected from 
the cluster. The period distribution of spun-up pulsars 
modeled under the assumption of accretion-induced 
NS magnetic field decay to a minimum of $10^8$ G is 
also in qualitative agreement with the observational 
data (Camillo and Rasio 2005). 

The situation with accreting NSs is more complex.
In our calculations, we found their number to 
be twice that of spun-up millisecond pulsars. The 
observed situation is opposite: there are a few lowmass 
X-ray binaries among the identified sources in 
the GC (Verbunt and Bassa 2004). Note, however, 
that recently the so-called quiescent low-mass X-ray 
binaries have been increasingly identified among 
weak X-ray sources in GCs with luminosities $L_x\sim 10^{31}-10^{32}$ erg s$^{-1}$ (see, e.g., Caskett et al. (2005) in 
Terzan 5 and Heinke et al.(2005) in 47 Tuc). The 
number of modeled binaries of this type will probably 
also decrease when we include the binary-binary 
interactions, which have not yet been included in 
our model. These types of interaction can reduce the 
binary fraction in GCs with time (for recent calculations, 
see Ivanova et al.2005). Indeed, when interacting 
with single stars, a binary system can be 
disrupted only if the energy  $E=mV_\infty^2/2$ of the impinging 
star exceeds the absolute value of the binding 
energy $|E_b|=G\frac{m_1m_2}{2a}$ for the binary. Here, $V_\infty$ is the 
relative velocity of the impinging star (infinitely far 
from the binary); $m_1, m_2$, and $m$ are the masses of 
the binary components and the impinging star; and $a$ 
is the semimajor axis of the binary. Clearly , for ''rigid''
binaries ($|E_b|>m\bar{V}^2/2,~ \bar{V}$ is the stellar velocity dispersion 
in the GC), the disruptions are more efficient 
when they interact with binary systems. Since we did 
not consider the interactions between binary systems 
in our calculations, the disruption rate of rigid binaries 
was definitely underestimated. 

The constructed X-ray luminosity function for accreting 
NSs in GCs has a mean slope, $d\log N/d\log L_x\sim -1.3$, close to the observed value for lowmass 
X-ray binaries in galactic bulges and elliptical 
galaxies. This universality of the luminosity function 
for X-ray sources pointed out by Gilfanov (2004) 
can be explained by common properties of the accretion 
in low-mass X-ray binaries (Postnov and Kuranov 
2005). 

Note that all of the numerical estimates in our 
calculations depend significantly on the assumed binary 
fraction in the GC. This parameter may be important 
for the evolution of the cluster as a whole 
(Fregeau et al.2005). Our calculations do not allow 
this parameter to be separated from other parameters 
(such as the mass range for stars with degenerate ONe-
Mg cores, the total number of stars in the GC, 
etc.). Nevertheless, reasonable agreement between 
our calculations and the observational data suggests 
that our main model assumptions about the evolution 
of stars in the GC are valid. 

Our description of the evolution of binary and 
single stars in GCs is not self-consistent, since it 
does not take into account some of the dynamical 
processes, such as binary-binary interactions; we 
also disregarded the reverse effect of binaries on the 
background of single stars by assuming the latter to 
be replenished to the previous level after each interaction.
These processes are important in calculating 
the formation of NS+NS pairs inGCs (Grindlay et al. 
2005). We plan to take into account these effects 
in subsequent papers. Nevertheless, the main result 
of our work will not change greatly: the hypothesis 
about the birth of some of the NSs in binary systems 
without any kick velocity allows us to obtain in a 
natural way a sufficient number of neutron stars in 
globular clusters without invoking an artificial retention 
factor, which becomes unavoidable for explaining 
the observations if any NS during its birth is assumed 
to acquire a high kick velocity. 

\Section{ACKNOWLEDGMENTS }
We wish to thank L.R. Yungelson for constructive 
remarks. This work was supported by the Russian 
Foundation for Basic Research (project nos.03-02- 
16110a and 03-02-06733mas).

\eject
\newpage
\newpage

\eject
\newpage


\pagebreak

\eject
\newpage
\eject
\newpage
\eject
\newpage
\newpage

\begin{table*}
\caption{GC model parameters used in our calculations. The central star density $n_c$ and the core radius $r_c$ are given at the core 
collapse time.}
\label{table} 
\begin{center}
\begin{tabular}{c c c c c c}
\hline  
 & Initial  & Initial  & Central  & Core     & Collapse \\
Model      & GC mass &  binary    &density   & radius  & time,\\
      &  M$_\odot$& fraction     &$n_c$ pc$^{-3}$ & $r_c$, pc   & Myr\\
\hline
A     & $5\times 10^5$ & 0.5     & $10^5$     & 0.3              & 7000\\
B     & $5\times 10^5$ & 0.5     & $10^6$     & 0.1              & 7000\\
\hline 
\end{tabular}
\end{center}
\end{table*}

\newpage
\eject
\newpage
\newpage

\unitlength 1mm

\begin{figure*}[ht]
\centering
\caption{Changes in cluster parameters: central density $\rho_c$ 
($pc^{-3}$), core radius $r_c$ (pc), and core star velocity dispersion 
$v_c$ (km s$^{-1}$ during the GC evolution. The GC core 
collapse time is indicated by the vertical line (7 Gyr).
\label{gc_evol}}
\end{figure*}

\begin{figure*}[ht]
\centering
\caption{Time evolution of the total number of binaries with neutron stars as a function of the type of secondary component: 
neutron stars in pairs withmain-sequence stars (NS + MS), neutron stars with white dwarfs (NS +WD), and binary neutron 
stars (NS + NS). The number of binaries inside the cluster (upper left panel) and those escaped from the cluster (upper right 
panel) for model A and the number of binaries insider the cluster (lower left panel) and those escaped from the cluster (lower 
right panel) formodel B are shown. The GC core collapse time is indicated by the vertical dotted line (7Gyr). 
\label{ns_types}}
\end{figure*}

\begin{figure*}[ht]
\centering
\caption{Formation rates of binaries with neutron stars for dynamical interactions between stars: the formation rates (Gyr$^{-1}$) 
in the case of exchange interactions between binary and single stars for model A (upper left panel) and model B (upper right 
panel) and the formation rates of binaries through tidal captures for model A (lower left panel) and model B (lower right panel) 
are shown. The GC core collapse time is 7 Gyr. 
\label{ns_accr}}
\end{figure*}

\begin{figure*}[ht]
\centering
\caption{Formation rates of low-mass X-ray binaries (LMXBs) and millisecond pulsars (MPSRs) in the case of interactions 
between stars: the formation rates (Gyr$^{-1}$) in the case of exchange interactions between binary and single stars for model A 
(upper left panel) and model B (upper right panel) and the formation rates of binaries through tidal captures for model A (lower 
left panel) and model B (lower right panel) are shown. The GC core collapse time is 7 Gyr. 
\label{ns_accr}}
\end{figure*}

\begin{figure*}[ht]
\centering
\caption{
Number of accreting neutron stars (as a function of the type of Roche-lobe-filling component: (Na + MSGW) in a pair 
with a main-sequence star (LMXBs) and (Na+WD) in a pair with a white dwarf and millisecond pulsars (single MPSRs in 
binaries). The numbers of systems inside the cluster (upper left panel) and those escaped from the cluster (upper right panel) 
for model A and the numbers of systems inside the cluster (lower left panel) and those escaped from the cluster (lower right 
panel) for model B are shown. The GC core collapse time is indicated by the vertical dotted line (7 Gyr). 
\label{ns_accr}}
\end{figure*}

\begin{figure*}[ht]
\centering
\caption{Left panel: ratios of the total number of millisecond pulsars (single + binary members) to the number of low-mass 
X-ray binaries in GC vs. time for models A (solid line) and B (dashed line). Right panel: ratios of the total number of neutron 
stars at the ejection phase (single + binary members) to the number of low-mass X-ray binaries in GC vs. time for models A 
(solid line) and B (dashed line).
\label{ratio}}
\end{figure*}

\begin{figure*}[ht]
\centering
\caption{
Fig. 7. Time evolution of the X-ray luminosity function for accreting neutron stars in GC formodels A (left panel) and B (right 
panel), in units of  $10^{38}$ erg s$^{-1}$.
\label{lumf}}
\end{figure*}

\begin{figure*}[ht]
\centering
\caption{ Histograms illustrating the time evolution of the spin period distribution for pulsars spun up previously at the accretion 
phase in GC for models A (left panel) and B (right panel).
\label{period_rec}}
\end{figure*}

\eject
\newpage

\end{document}